\def\type{arxiv}

\def\A{arxiv}
\def\J{journal}

\ifx\type\A
\documentclass[12pt]{amsart}
\else
\documentclass{ap-jnmp}
\fi

\usepackage{mathrsfs}

\DeclareMathOperator{\diag}{diag}

\newcommand{\BigFig}[1]{\parbox{12pt}{\Huge #1}}
\newcommand{\BigZero}{\BigFig{0}}
\begin{document}

\ifx\type\J
%\arttype{Letter} % default 'Article'

\markboth{Pasha Zusmanovich}{On near and the nearest correlation matrix}

\catchline{}{}{}{}{}
%
%%%%%%%%%%%%%%%%%%%%%%%%%%%%%%%%%%%%%%%%%%%%%%%%%%%%%%%%%%%%%%%%%%%%
\copyrightauthor{P. Zusmanovich}

\title{On near and the nearest correlation matrix}
\author{\footnotesize Pasha Zusmanovich\footnote{
Current address: 
Department of Mathematics and Physics, North Carolina Central University, 
1801 Fayetteville St., Durham, NC 27707, USA
}}
\address{
Department of Mathematics, Tallinn University of Technology, Ehitajate tee 5, 
Tallinn 19086, Estonia 
\\
\email{pasha.zusmanovich@ttu.ee}
}
\maketitle
\thispagestyle{empty} 

\else

\title{On near and the nearest correlation matrix}
\author{Pasha Zusmanovich}
\address{
Department of Mathematics, Tallinn University of Technology, Ehitajate tee 5, 
Tallinn 19086, Estonia
}
\curraddr{
Department of Mathematics and Physics, 
North Carolina Central University, 1801 Fayetteville St., Durham, NC 27707, USA
}
\email{pzusmanovich@nccu.edu}
\date{Last revised August 29, 2013}
\thanks{\textsf{arXiv:1303.3226}}

\fi

\ifx\type\J
\vphantom{\vbox{
\begin{history}
\received{(28 July 2012)}
\revised{(29 August 2012)}
%\accepted{(14 March 2009)}
%\comby{(xxxxxxxxx)}
\end{history}
}}
\fi

\begin{abstract}
We present an elementary heuristic reasoning based on Arnold's theory of 
versal deformations in support of a straightforward algorithm for 
finding a correlation matrix near a given symmetric one. 
\end{abstract}


\ifx\type\J

\ccode{2010 Mathematics Subject Classification: 
14D99, 15A18, 15A21, 15B48, 62P05, 62P35, 65F35}

\else

\maketitle

\fi

\section*{Introduction}

Bankers are interested in correlations between time series associated with
various financial instruments 
(such as prices of stocks, options, futures and other derivatives, 
currency exchange rates, etc.),
presented in the form of the sample correlation matrix. 
As a \emph{bona fide} correlation matrix, it should be positive semidefinite. 
In practice, however, the computed matrix almost always turns out to be not 
positive semidefinite. The main reason for this is twofold:
methodological errors (taking data for different instruments in 
different time ranges, inconsistent approach to inventing missing data), and
floating point rounding errors.

The computed correlation matrix is utilized, however, in further analysis, like 
evaluation of various risks; for this, its positive semidefiniteness is 
crucial. As in the most cases it is impossible to backtrack the 
origin of the problem due to shortage of time, the large amount of numerical 
data (a typical scenario may involve daily computed correlation matrices reaching
the size of ten thousand by ten thousand), 
and complexity of the methods used in its retrieval, processing and 
storage, one usually resorts on ``correcting'' the symmetric matrix at hand
to make it positive semidefinite.

Naturally, this ``correction'' should be as small as possible.
So, a practical problem arises: for a given symmetric matrix, find the nearest,
in some sense, correlation matrix. 
A quick glance at the literature (mentioned below) suggests that this problem 
arises not only in banking.

Not surprising then that this problem attracted a considerable attention.
While no exact expression for the nearest correlation matrix is available, 
many papers -- see \cite{higham}, \cite{qxx} and references therein --
contain algorithms for its determination. These algorithms utilize methods from
convex analysis, semismooth optimization, and other sophisticated branches of 
numerical mathematics.
Earlier results in this direction are also surveyed in \cite[\S 9.4.6]{gentle}.
In all these works, ``nearest'' is understood in the sense of the Frobenius 
matrix norm, or some its (weighted) variation.

In the real life, however, bankers tend to ignore all this wisdom and 
implement a very pedestrian approach to this problem (sometimes called
``shrinking'' and which can be found, with some variations, 
in \cite{sandia}, \cite{qxx}, \cite{primitive}, \cite[Exercise 9.14]{gentle}, 
and in many other places). Namely, in the spectral decomposition 
$A = B J B^\top$ of a given $n \times n$ symmetric matrix $A$, 
where $B$ is an orthogonal matrix of eigenvectors, and
\begin{equation}\label{spectral}
J =
\left(
\begin{matrix}
\lambda_1 & &           &                   \\
            & \lambda_2 &        & \BigZero \\
\BigZero    &           & \ddots &          \\
            &                    &        & \lambda_n
\end{matrix}
\right)
\end{equation}
is a diagonal matrix of eigenvalues of $A$, replace all negative eigenvalues 
by some small positive number $\varepsilon$: 
$$
\widehat\lambda_i = \begin{cases}
\varepsilon &\text { if } \lambda_i < 0    \\
\lambda_i   &\text { if } \lambda_i > 0 ,
\end{cases}
$$ 
for $i=1,\dots,n$ (in practice, zero eigenvalues do not occur). 
The resulting matrix 
$$
(\widehat{a}_{ij})_{i,j=1}^n = 
B 
\left(
\begin{matrix}
\widehat\lambda_1 & &                 &                   \\
                  & \widehat\lambda_2 &        & \BigZero \\
\BigZero          &                   & \ddots &          \\
                  &                   &        & \widehat\lambda_n
\end{matrix}
\right)
B^\top
$$
is a positive definite covariance matrix, and its normalization 
\begin{equation}\label{normaliz}
\Big(
\frac{\widehat a_{ij}}{\sqrt{\widehat a_{ii} \widehat a_{jj}}}
\Big)_{i,j=1}^n
\end{equation}
is declared to be the requested correlation matrix, allegedly close to the initial matrix 
$A$.

This pedestrian approach turns out to be very efficient in practice
(in all banking numerical examples we have observed, the initial and corrected 
matrices were very close with respect to the max norm\footnote[2]{
In \S \ref{proof}, at a certain place we use submultiplicativity (i.e.,
$\| A B \| \le \| A \| \| B \|$ for any two matrices $A$, $B$) of the matrix 
norm $\| \cdot \|$ measuring the ``nearness''.
The max norm is not submultiplicative, so, formally, it does
not fit those arguments. This can be remedied, however, by a minor 
(and well-known) fix: the max norm becomes submultiplicative when
multiplied by the matrix size (see, for example, \cite[p.~292]{horn-j}). 
Even taking into account this factor ($<10^5$ in practice), the absolute values
of differences between the corresponding elements of the initial and corrected 
matrices remained very small in all real-life examples we have seen.}, 
and no discrepancies occurred utilizing the corrected matrix in
the subsequent analysis).
In this note we offer an heuristic argument explaining this, perhaps, 
unreasonable at the first glance, efficiency.
The argument, presented in \S \ref{proof}, is an easy application of 
Arnold's theory of versal deformation of matrices. A fragment of the
theory needed for our purposes is briefly recalled in \S \ref{sec-versal}.
The last \S 3 contains an example.

\section{Arnold's theory of versal deformations}\label{sec-versal}

In 1971, Vladimir Arnold developed a theory of versal deformations of matrices, 
which triggered a wake of subsequent work. The original paper \cite{arnold} is 
still the best exposition of this theory. The main result of this theory can be
formulated in many different ways, one of them runs as follows.

Let $A$ be a complex $n \times n$ matrix with distinct eigenvalues 
$\lambda_1, \lambda_2, \dots, \lambda_k$, and with the Jordan normal form
\begin{equation*}
J =
\left(
\begin{matrix}
                      J^{\lambda_1}_{n_{11}, \dots, n_{1m_1}} & & &        \\
         &            J^{\lambda_2}_{n_{21}, \dots, n_{2m_2}} & & \BigZero \\
\BigZero & & \ddots &                                                      \\
         & &        & J^{\lambda_k}_{n_{k1}, \dots, n_{km_k}}     
\end{matrix}
\right)
\end{equation*}
where 
\begin{equation*}
J^\lambda_\ell =
\left(
\begin{matrix}
\lambda & 1        &        &          &         \\
        & \lambda  &        & \BigZero &         \\
        &          & \ddots &          &         \\
        & \BigZero &        & \lambda  & 1       \\
        &          &        &          & \lambda 
\end{matrix}
\right)
\end{equation*}
is one Jordan block of size $\ell\times\ell$, and
\begin{equation*}
J^\lambda_{n_1, \dots, n_m} =
\left(
\begin{matrix}
J^\lambda_{n_1} &                 &        &          \\
                & J^\lambda_{n_2} &        & \BigZero \\
\BigZero        &                 & \ddots &          \\
                &                 &        & J^\lambda_{n_m}     
\end{matrix}
\right)
\end{equation*}
consists of all Jordan blocks of sizes
$n_1 \times n_1, n_2 \times n_2, \dots, n_m \times n_m$, arranged in the 
non-increasing order (i.e., $n_1 \ge n_2 \ge \dots \ge n_m$), 
corresponding to a single eigenvalue $\lambda$ with algebraic multiplicity
$n_1 + n_2 + \dots + n_m$.

Let us define a parametric deformation 
\begin{equation*}
J(\xi_{11}, \xi_{12}, \dots, \xi_{1N_1}, \xi_{21}, \xi_{22}, \dots, \xi_{2N_2},
\dots, \xi_{k1}, \xi_{k2}, \dots, \xi_{kN_k})
\end{equation*}
of $J$, with complex parameters $\xi_{11}, \dots, \xi_{kN_k}$, where 
$$
N_i = n_{i1} + 3n_{i2} + 5n_{i3} + \dots + (2m_i-1)n_{im_i} 
$$
for $i = 1, \dots, k$.

First, all blocks corresponding to different eigenvalues are deformed 
independently:
\begin{equation*}
\left(
\begin{matrix}
J^{\lambda_1}_{n_{11}, \dots, n_{1m_1}}(\xi_{11}, \dots, \xi_{1N_1}) & & &   \\
&J^{\lambda_2}_{n_{21}, \dots, n_{2m_2}}(\xi_{21}, \dots, \xi_{2N_2})&&\BigZero \\
\BigZero & & \ddots &                                                      \\
& & & J^{\lambda_k}_{n_{k1}, \dots, n_{km_k}}(\xi_{k1}, \dots, \xi_{kN_k})     
\end{matrix}
\right) .
\end{equation*}
Second, a single Jordan block $J^\lambda_\ell$ is deformed as follows:
\begin{equation*}
J^\lambda_\ell(\chi_1, \dots, \chi_l) =
\left(
\begin{matrix}
\lambda & 1       &        &            & \BigZero        \\
0       & \lambda &        &            &                 \\
\vdots  & \vdots  & \ddots &            &                 \\
0       & 0       & \dots  & \lambda    & 1               \\
\chi_1  & \chi_2  & \dots  & \chi_{l-1} & \lambda +\chi_l 
\end{matrix}
\right)
\end{equation*}
and, finally, the deformation 
$J^\lambda_{n_1, n_2, \dots, n_m}
(\chi_1, \dots, \chi_{n_1 + 3n_2 + \dots +(2m-1)n_m})$ of
all blocks corresponding to a single eigenvalue $\lambda$
is defined in the following recursive way:
\begin{equation*}
\left( \hspace{-\arraycolsep}
\begin{array}{cccccccccc}
\lambda&1                                                                      \\
0      &\lambda&       &            &                  &            &&\BigZero         \\
\vdots &\vdots &\ddots &            &                  &            && \\
0&0    &\dots  &\lambda&1                                                      \\
\chi_1 &\chi_2 &\dots  &\chi_{n_1-1}&\lambda+\chi_{n_1}&\chi_{n_1+1}&\chi_{n_1+2}&\dots&\chi_{n_1+n_2+\cdots+n_m-1}&\chi_{n_1+n_2+\cdots+n_m} \\
\multicolumn{1}{l}{\chi_{n_1+n_2+\cdots+n_m+1}}   \\
\multicolumn{1}{l}{\chi_{n_1+n_2+\cdots+n_m+2}}   \\
\vdots &       &\BigZero&&&\multicolumn{5}{c}{J^\lambda_{n_2,\dots,n_m}\big(\chi_{n_1+2n_2+\cdots+2n_m+1},\dots,\chi_{n_1+3n_2+\dots+(2m-1)n_m}\big)} \\
\multicolumn{1}{l}{\chi_{n_1+2n_2+\cdots+2n_m-1}} \\
\multicolumn{1}{l}{\chi_{n_1+2n_2+\cdots+2n_m}} 
\end{array}
\hspace{-\arraycolsep} \right) .
\end{equation*}

Then, according to \cite[Theorem 4.4]{arnold}, 
any smooth family of complex $n \times n$ matrices containing $A$, and 
parametrized by several complex variables $\mathbf{t} = (t_1, t_2, \dots)$, 
can be represented, in a sufficiently small neighborhood of 
$\mathbf{0} = (0,0,\dots)$, as the
product
\begin{equation}\label{versal}
B(\xi_1(\mathbf{t}), \dots, \xi_N(\mathbf{t})) \> 
J(\xi_1(\mathbf{t}), \dots, \xi_N(\mathbf{t})) \>
B(\xi_1(\mathbf{t}), \dots, \xi_N(\mathbf{t}))^{-1} ,
\end{equation}
where $N = \sum_{i=1}^k N_i$, all $\xi_i$'s are smooth functions of their 
arguments vanishing at $\mathbf{0}$, $B(\xi_1, \dots, \xi_N)$ is a smooth family
of invertible matrices, and 
$$
A = A(\mathbf{0}) = B(\mathbf{0}) \> J(\mathbf{0}) \> B(\mathbf{0})^{-1} .
$$

If the spectrum of $A$ is simple, then this picture is significantly 
streamlined.
The total number $N$ of parameterizing functions in (\ref{versal}) is equal to 
$n$, the size of the matrix, and the deformation family of the diagonal matrix
(\ref{spectral}) itself consists of diagonal matrices:
\begin{equation*}
J(\xi_1(\mathbf{t}), \dots, \xi_n(\mathbf{t})) =
\left(
\begin{matrix}
\lambda_1 + \xi_1(\mathbf{t}) &                   &        &          \\
                  & \lambda_2 + \xi_2(\mathbf{t}) &        & \BigZero \\
\BigZero          &                   & \ddots &          \\
                  &                   &        & \lambda_n + \xi_n(\mathbf{t})
\end{matrix}
\right)
\end{equation*}
where $\xi_i(\mathbf{0}) = 0$, $i=1,\dots,n$.

There are corresponding results for matrices with real coefficients \cite{galin}
and symmetric matrices \cite{patera} (as well as for many other situations
in which a Lie group acts on a manifold, see \cite{arprob}), 
which are technically more complicated.
However, for our purpose it suffices to use Arnold's original
setting. Just note that as we are interested solely in symmetric matrices which
are brought to the diagonal form (\ref{spectral}) by an orthogonal 
transformation, the combination
of results of \cite{arnold} and \cite{patera} shows that in the decomposition 
(\ref{versal}) we may assume that all matrices in the family $B$ are orthogonal,
i.e.
$$
B(\xi_1(\mathbf{t}), \dots, \xi_N(\mathbf{t}))^{-1} =
B(\xi_1(\mathbf{t}), \dots, \xi_N(\mathbf{t}))^\top
$$
for all $\mathbf{t}$ from an appropriate neighborhood of zero.

Arnold's theory can be considered as one of the instances of the 
algebro-geometric deformation theory of various kinds of algebraic objects, 
along with deformation theory of algebras, of modules over algebras, and of morphisms
between algebras, see \cite{fialowski-oh} for overview of those theories from
a unifying viewpoint.

\section{Just getting rid of negative eigenvalues is enough}\label{proof}

It is well-known that the set of the correlation matrices coincides 
with the set of (real) positive semidefinite matrices with units on the main 
diagonal (see, for example, \cite[Chapter III, \S 6, Theorem 4]{feller}), 
so we will use these two notions interchangeably.

Suppose $A$ is a symmetric $n\times n$ matrix with units on the main diagonal. 
As the set of matrices
with simple spectrum is Zariski-dense in the set of all real $n\times n$ 
matrices, we may assume that $A$ has simple spectrum (a more down-to-earth 
incarnation of this fact is that all correlation matrices appearing in banking
practice, and, more generally, correlation matrices based on a sufficiently 
large amount of real-world data, have simple spectrum; in fact, the reasonings
below could be modified for the case of arbitrary spectrum, but technically
they would become more complicated). Let (\ref{spectral}) be its Jordan normal 
form, all $\lambda_i$'s being pairwise distinct (and some of them are negative,
of small absolute value).

Suppose further that there exists a correlation matrix $C$ ``near'' $A$,
and that $A$ and $C$ are members of a smooth family of matrices. The latter
assumption is justified both from theoretical (correlation matrix is a smooth
function of time series it correlates between) and practical 
(the financial processes a correlation matrix is trying to capture, are assumed
to be satisfactorily modelled by smooth functions) viewpoints.

According to the theory presented in \S \ref{sec-versal}, in a sufficiently
small neighborhood $\mathscr U$ of $A = A(\mathbf{0})$, we may write this smooth
family in the following parametric form:
\begin{equation}\label{yoyo}
A(\mathbf{t}) = B(\mathbf{t}) 
\left(
\begin{matrix}
\lambda_1 + \xi_1(\mathbf{t}) &                      &        &          \\
                     & \lambda_2 + \xi_2(\mathbf{t}) &        & \BigZero \\
\BigZero             &                      & \ddots &                   \\
                     &                      &        &\lambda_n+\xi_n(\mathbf{t})
\end{matrix}
\right)
B(\mathbf{t})^\top
\end{equation}
for some smooth functions $\xi_i$ such that $\xi_i(\mathbf{0}) = 0$ for all 
$i = 1, \dots, n$, and a smooth family 
$B(\mathbf{t}) = \big(b_{ij}(\mathbf{t})\big)_{i,j=1}^n$ of orthogonal matrices. 
In particular, $C$, being a member of the family, is represented in the form
(\ref{yoyo}) for some value $\mathbf{t} = \mathbf{t}_0$.

The condition of positive definiteness of a member of the family $A(\mathbf{t})$
is equivalent to 
\begin{equation}\label{lambda}
\xi_i(\mathbf{t}) > -\lambda_i
\end{equation}
for all $i=1,\dots,n$, and the condition of having units on the main diagonal is 
equivalent to
\begin{equation}\label{sol}
B^{\circ 2}(\mathbf{t})
\left(
\begin{matrix}
\lambda_1 + \xi_1(\mathbf{t}) \\
\lambda_2 + \xi_2(\mathbf{t}) \\
\vdots                        \\
\lambda_n + \xi_n(\mathbf{t})
\end{matrix}
\right) = 
\left(
\begin{matrix}
1      \\
1      \\
\vdots \\
1
\end{matrix}
\right) ,
\end{equation}
where $B^{\circ 2}(\mathbf{t}) = \big(b_{ij}(\mathbf{t})^2\big)_{i,j=1}^n$ is the
Hadamard square of $B(\mathbf{t})$.
The set of solutions of (\ref{sol}) contains at least two points, 
$\mathbf{0}$ and $\mathbf{t}_0$, hence it forms a nonempty variety
$\mathscr H$ in the space of parameters, and the intersection of this 
variety with the neighborhood $\mathscr U$ and the open domain defined by 
conditions (\ref{lambda}), defines a certain neighborhood $\mathscr U^\prime$
of $C = A(\mathbf{t}_0)$ in $\mathscr H$.

In terms of the procedure described in the introduction, getting $C$
from $A$ amounts to ``adjusting'' eigenvalues, i.e., adding to each eigenvalue
$\lambda_i$ in the diagonal form (\ref{spectral}) a small correction 
$\xi_i(\mathbf{t}_0)$, and subsequent ``normalization'' (\ref{normaliz}); all 
this corresponds to getting back correlation matrix 
$$
B(\mathbf{t}_0) 
\left(
\begin{matrix}
\lambda_1 + \xi_1(\mathbf{t}_0) &                      &        &          \\
                     & \lambda_2 + \xi_2(\mathbf{t}_0) &        & \BigZero \\
\BigZero             &                      & \ddots &                   \\
                     &                      &        &\lambda_n+\xi_n(\mathbf{t}_0)
\end{matrix}
\right)
B(\mathbf{t}_0)^\top
$$
from the covariance matrix 
$$
B(\mathbf{0}) 
\left(
\begin{matrix}
\lambda_1 + \xi_1(\mathbf{t}_0) &                      &        &          \\
                     & \lambda_2 + \xi_2(\mathbf{t}_0) &        & \BigZero \\
\BigZero             &                      & \ddots &                   \\
                     &                      &        &\lambda_n+\xi_n(\mathbf{t}_0)
\end{matrix}
\right)
B(\mathbf{0})^\top .
$$

Assuming that the neighborhood $\mathscr U^\prime$ is small enough, any matrix 
from it will do, but what will be the best choice? As mentioned in the 
introduction, this is, generally, a difficult problem not admitting a 
closed-form solution. Intuitively, there is no need to adjust the positive 
eigenvalues, but only the negative ones, and the following imprecise reasoning 
supports this.

Assuming that the matrix norm $\| \cdot \|$ measuring the ``nearness'' is 
submultiplicative and is invariant under transposition (the latter assumption is 
not essential but slightly simplifies the expressions below), we have:
\begin{alignat}{1}\label{ineq}
\| A(\mathbf{t}) - A(\mathbf{0}&) \| 
\notag \\
=& \phantom{2}
\| 
B(\mathbf{t}) \diag\big(\lambda_1 + \xi_1(\mathbf{t}),\dots,\lambda_n + \xi_n(\mathbf{t})\big)
B(\mathbf{t})^\top - B(\mathbf{0}) \diag\big(\lambda_1,\dots,\lambda_n\big)
B(\mathbf{0})^\top 
\| 
\notag \\
\le& \phantom{2} \| \diag\big(\xi_1(\mathbf{t}),\dots,\xi_n(\mathbf{t})\big) \|
                 \| B(\mathbf{t}) - B(\mathbf{0}) \|^2
\notag \\
+& 2 \| \diag\big(\xi_1(\mathbf{t}),\dots,\xi_n(\mathbf{t})\big) \|
     \| B(\mathbf{t}) - B(\mathbf{0}) \| \| B(\mathbf{0}) \| 
\\
+& \phantom{2} \| \diag\big(\xi_1(\mathbf{t}),\dots,\xi_n(\mathbf{t})\big) \|
               \| B(\mathbf{0}) \|^2
\notag \\
+& \phantom{2} \| B(\mathbf{t}) - B(\mathbf{0}) \|^2 
               \| \diag\big(\lambda_1,\dots,\lambda_n\big) \| 
\notag \\
+& 2 \| B(\mathbf{t}) - B(\mathbf{0}) \| \|B(\mathbf{0}) \| 
     \| \diag\big(\lambda_1,\dots,\lambda_n\big) \| .
\notag
\end{alignat}
Both theoretical considerations in \cite{arnold}, and computational procedures 
developed in \cite{mailybaev} suggest that matrix entries of the parametric 
family $B(\mathbf{t})$ providing the transformation to the canonical form 
(\ref{yoyo}) of the versal deformation, have, as power series of the parameter 
$\mathbf{t}$, the same order of magnitude as matrix entries of the canonical 
form itself. In particular, in a sufficiently small neighborhood of zero, which 
can be assumed lying inside $\mathscr U$, we have
$$
       \| B(\mathbf{t}) - B(\mathbf{0}) \| \le 
\alpha \| \diag\big(\xi_1(\mathbf{t}),\dots,\xi_n(\mathbf{t})\big) \|
$$
for some (positive) constant $\alpha$. This, together with (\ref{ineq}), implies
that $\| A(\mathbf{t}) - A(\mathbf{0}) \|$ is bounded by a cubic polynomial in  
$\| \diag\big(\xi_1(\mathbf{t}),\dots,\xi_n(\mathbf{t})\big) \|$ with 
positive coefficients. The latter polynomial is a monotonic function, so to 
minimize $\| A(\mathbf{t}) - A(\mathbf{0}) \|$ one may wish to minimize
$\| \diag\big(\xi_1(\mathbf{t}),\dots,\xi_n(\mathbf{t})\big) \|$ instead.
Subject to restriction (\ref{lambda}), for all matrix norms appearing 
in practice, this amounts to setting $\xi_i(\mathbf{t})$ to a positive value 
``just a little bit'' bigger than $-\lambda_i$ if $\lambda_i$ is negative, and 
to zero otherwise.

\bigskip

We emphasize that these are merely non-rigorous, heuristic, arguments,
and by no means they can substitute a rigorous analysis given in 
\cite{higham}, \cite{qxx} and similar papers.
However, these arguments perfectly suit the practical nature of the problem: 
one knows \textit{a priori} that a very close correlation matrix exists.
In such a situation, Arnold's theory guarantees existence of
such matrix in the simple form (\ref{yoyo}). Though it is not guaranteed that 
this will be the \textit{nearest} correlation matrix, it certainly will be
a \textit{near} one, and this suffices in practice.

Of course, arguments of this sort can be used in other similar 
situations -- for example, to justify adjusting (``cutoff'') of some unwanted, 
from the physical perspective, eigenvalues of (valid) correlation matrices 
arising in lattice gauge theory (see \cite{yjjl} and references therein),
or correcting the degenerate covariance matrix from an insufficient
amount of data in the situation when the number of observations is much smaller
then the number of variables (see \cite{tw} and references therein).

\section{An example}

Here we present a ``toy'' example illustrating the procedure described above.
Being a toy one, this example, however, adequately reflects what is happening in 
the ``real life'' in banks. Other examples may be found in \cite{sandia} and  
\cite{primitive}.

Let us take the close price of 4 stocks traded at 
Euronext Amsterdam: 
Galapagos, Heineken, TomTom, Wolters Kluwer, as well as Euro/US dollar 
rate, for 6 consecutive business days during the period 
from July 26 till August 2, 2013 
(available at the time of writing at \texttt{http://www.aex.nl/}):

\smallskip

\begin{center}
{\small
\begin{tabular}{|l|r|r|r|r|r|r|}
\hline
& Jul 26 & Jul 29 & Jul 30 & Jul 31 & Aug 1 & Aug 2 \\ 
\hline 
Galapagos      & 16.08\phantom{0} & 16.15\phantom{0} & 16.13\phantom{0} & 
                 16.25\phantom{0} & 16.25\phantom{0} & 16.23  \\ \hline 
Heineken       & 50.69\phantom{0} & 50.88\phantom{0} & 51.66\phantom{0} & 
                 52.8\phantom{00} & 53.8\phantom{00} & 53.9\phantom{00} \\ \hline
TomTom         & 4.286            & 4.3\phantom{00}  & 4.363            & 
               4.38\phantom{0}    & 4.497            & 4.525            \\ \hline
Wolters Kluwer & 18.09\phantom{0} & 18.005           & 18.095           & 
                 18.145           & 18.5\phantom{00} & 18.515           \\ \hline
Euro/US dollar & 1.3279 & 1.3263 & 1.3266 & 1.3300 & 1.3212 & 1.328     \\   
\hline
\end{tabular}
}
\end{center}

\medskip

Let us compute now the corresponding correlation matrix. Assume, however, that,
when computing correlation between Wolters Kluwer and 
Euro/US dollar, the last day data for one of these instruments was lost, and
the correlation was computed for the first 5 days only. That leads to the 
following ``distorted'' correlation matrix:
$$
\left(
\begin{matrix}
1 & 0.896 & 0.785 & 0.684 & -0.179 \\
  & 1     & 0.970 & 0.914 & -0.275 \\
  &       & 1     & 0.961 & -0.359 \\
  &       &       & 1     & -0.767 \\
  &       &       &       &  1         
\end{matrix}
\right) .
$$
This matrix, unlike the ``real'' correlation matrix, is not positive definite, 
with the smallest eigenvalue equal to $-0.089$.

Applying the procedure described in the Introduction to this distorted matrix, 
with $\varepsilon = 0.001$, we get a positive definite matrix
$$
\left(
\begin{matrix}
1 & 0.888 &  0.779 & 0.672 & -0.175 \\
  & 1     &  0.970 & 0.860 & -0.284 \\
  &       &  1     & 0.909 & -0.366 \\
  &       &        & 1     & -0.716 \\
  &       &        &       &  1
\end{matrix}
\right) ,
$$
which is ``close enough'' to the distorted one.

\section*{Acknowledgements}

This note was essentially written around 2005, during my employment at ING Bank.
Thanks are due to Manon ten Voorde, who, in those days, introduced me to 
the problem of finding the nearest correlation matrix and explained its 
peculiarities. During the final write-up I was supported by grants ERMOS7 
(Estonian Science Foundation and Marie Curie Actions) and ETF9038 
(Estonian Science Foundation).

\end{document}